\documentstyle[epsfig,12pt]{article}
\begin{document}

\newcommand \be  {\begin{equation}}
\newcommand \bea {\begin{eqnarray} \nonumber }
\newcommand \ee  {\end{equation}}
\newcommand \eea {\end{eqnarray}}

\title{{\bf Rejuvenation in the Random Energy Model}}

\author{Marta Sales$^{1,2}$, Jean-Philippe Bouchaud$^{2}$}

\date{\it
$^1$ Dept. de F\'{\i}sica Fonamental, \\
Diagonal 647. Pta. 3a. \\
08028 Barcelona, Spain \\
$^2$ Service de Physique de l'\'Etat Condens\'e,\\
 Centre d'\'etudes de Saclay, \\ Orme des Merisiers, 
91191 Gif-sur-Yvette Cedex, France 
}
\maketitle

\begin{abstract}
We show that the Random Energy Model has interesting rejuvenation 
properties
in its frozen phase. Different `susceptibilities' to 
temperature changes, for the free-energy and
for other (`magnetic') observables, can be computed exactly. 
These susceptibilities diverge at the 
transition temperature, as $(1-T/T_c)^{-3}$ for the free-energy.
\end{abstract}

\vskip 1cm

A small temperature change in the low temperature phase of spin-glasses 
is able to
`rejuvenate' an already aged system \cite{RejMemory,StAndrews,Lengths}. 
More precisely, the a.c. susceptibility of a 
system aged for a very long time at $T_1 < T_c$ (where $T_c$ is the spin-glass phase
transition temperature) and then suddenly cooled at $T_1 - \Delta T$ is, provided $\Delta T$ is not too
small, very close the the susceptibility of a young system directly cooled from high 
temperatures to $T_1 - \Delta T$. This `fragility' to temperature changes has been interpreted 
early on as a signature of `temperature chaos', that is the fact that the equilibrium states 
of a disordered system are very different for different temperatures: beyond a certain 
length scale $\ell_{\Delta T}$ (which diverges for small $\Delta T$), the thermodynamical
states become uncorrelated and the overlap between them tends to zero for large system 
sizes. Such an effect was
conjectured in the context of the droplet model of spin-glasses, based on scaling arguments \cite{FH,KH},
and supported by Migdal-Kadanoff renormalisation group calculations \cite{BM}. A similar effect is 
also predicted for {\it pinned interfaces} \cite{FH2}, and has recently been checked in careful large scale
simulations and analytical arguments in the case of the $1+1$ directed polymer in random 
media \cite{Yoshino/Sales}. However, temperature chaos has been recently disproven 
in the mean-field SK model \cite{Rizzo}, and
not been found either in numerical simulations of the 3D Edwards-Anderson spin-glass model \cite{Billoire,
Ritort}.
This might be due to the fact that the length $\ell_{\Delta T}$ (if it exists) involves a large 
numerical 
prefactor and is therefore larger than numerically accessible sizes. It could also be that 
although the length scale $\ell_{\Delta T}$ plays a relevant role in the overlap between the
states at $T_1$ and $T_1 -\Delta T$, this overlap does tend very slowly to zero
for large sizes, as in \cite{Yoshino/Sales}. Finally, the `temperature chaos' interpretation of the 
rejuvenation effect has
to be compatible with the simultaneous {\it memory} that one observes experimentally. A scenario
for this was recently proposed in \cite{YLB}.

Another line of thought to explain rejuvenation and memory in spin-glasses is based on 
`hierarchical' energy landscape pictures \cite{Vincent,LeFloch}. Lower temperatures reveal finer 
details (rejuvenation) 
while large scale jumps are frozen out (memory) \cite{Bouchaud/Dean,StAndrews,Lengths}. Numerical 
simulations of 
the dynamics on Parisi's 
tree \cite{Nemoto/Sasaki} or in the Sinai potential \cite{Sales} indeed confirm 
that these effects exist in the 
absence of `true' chaos.
Even a simple two-level system does actually lead to
some rejuvenation when the temperature is of the order of the energy difference between the 
levels -- simply because the relative Boltzmann weight changes when the temperature is changed.
In this paper, we want to study in details the temperature rejuvenation effect in the glassy phase
of the Random Energy Model ({\sc rem}), for which a number of exact results are known \cite{Derrida}. This low
temperature phase is described, in the replica language, by a `one-step' replica symmetry 
breaking scheme \cite{Gross/Mezard}. This model is expected to be in the same universality class (with possibly 
minor corrections) as a large 
number of other models \cite{Revus}, such as the p-spin models (advocated to be good models for glasses), 
the Bernasconi model, the directed polymer (or random manifold)
in high dimensions \cite{Derrida/Spohn,Mezard/Parisi}, the unbinding transition of heteropolymers 
\cite{Tang/Chate} and  
the problem of a single particle in a logarithmically 
correlated random potential \cite{Carpentier/LeDoussal}. Interestingly, 
a dynamical version of the {\sc rem} naturally leads 
to aging dynamics \cite{Bouchaud/Dean,Monthus/Bouchaud,Maass}. 

Here, we want to compute exactly the `rejuvenation susceptibility' of different 
observables to small
temperature changes. Even if the model is not `chaotic', it reveals a number of interesting features 
that may be relevant to the present discussion. For example, the `susceptibility' to temperature
changes diverges when the temperature tends to the glass temperature of the model. Experimental   
consequences are discussed in the conclusion.

It is well known that the low temperature phase of the {\sc rem} is equivalent to that of the `trap' 
model \cite{Derrida3,Bouchaud/Mezard}, where $M$ energy states $\epsilon_i, \ i=1,...M$ are chosen with an exponential probability
distribution:
\be
P(\epsilon) = \frac{1}{T_c} \exp (-\frac{|\epsilon|}{T_c}).
\ee 
Note that $\epsilon$ is chosen to be negative. The partition function for this model is simply
$Z(T)=\sum_{i=1}^M z_i$ with $z_i=\exp(|\epsilon|_i/T)$. 
This model undergoes a phase transition at $T=T_c$,
where the partition function `localizes' on a few states. More precisely, in the limit $M \to 
\infty$, the Boltzmann weights of
a finite number of states add up to a finite fraction of the partition function for $T < T_c$
\cite{Derrida3,Bouchaud/Mezard}. 
Aging is the dynamical counterpart of this localization effect: most of the elapsed time is spent 
by the system in the deepest available well \cite{Bouchaud/Dean}.

As a first definition of the susceptibility to temperature changes, we study, following
Fisher and Huse \cite{FH2}, the 
correlation of the free-energy fluctuations for two different temperatures. More 
precisely, we write:
\be
C_F(T_1,T_2)= \frac{\overline{(\ell_1-\overline{\ell_1})
(\ell_2-\overline{\ell_2})}}
{\left( 
\overline{(\ell_1-\overline{\ell_1})^2}
\ \overline{(\ell_2-\overline{\ell_2})^2}\right)^{1/2}}.
\ee
where $\ell_1$ stands for $\log Z(T_1)$ and the overline means that 
we average over the distribution of the energies $\epsilon$.
When $T_1=T_2 (1+\varepsilon)$ we expect that:
\be
C_F(T_1,T_2)=1 - \kappa_F \varepsilon^2, \quad(\varepsilon \to 0),
\ee
where $\kappa_F$ defines the susceptibility to temperature changes. The calculation 
of this quantity starts with Derrida's representation of $\log Z$ \cite{Derrida}:
\be
\log Z = \int_0^\infty dt \ \frac{\exp(-t)-\exp(-t Z)}{t} = \lim_{b \to 0^+}
\int_0^\infty dt \ t^{b-1} \left(\exp(-t)-\exp(-t Z)\right)
\ee
where $b$ has been introduced to ensure convergence of intermediate calculation 
steps. The average over $\epsilon$ then involves:
\be
\overline{\exp -tZ}=(\overline{\exp -tz})^M=\left(1-(1-\overline{\exp -tz})\right)^M
\ee
For large $M$, only the vicinity of $t=0$ will therefore be of importance. 
Using the fact that 
the random variables $z$ are distributed with a power law-tail $\mu z^{-1-\mu}$ with
$\mu=T/T_c$, one finds that:
\be
1-\overline{\exp -tz} \sim_{t \to 0} \Gamma(1-\mu)t^\mu.
\ee
Using the last result in the previous two equations finally leads to the following 
result for the free-energy (in units of $T_c$): 
\be
-\mu \ \overline{\log Z}= -\log M - \gamma (1-\mu) - \log \Gamma(1-\mu),
\ee
where $\gamma$ is Euler's constant.
The first term comes from the fact that when $M$ is large, the smallest energy 
drawn from an exponential distribution behaves as $-T_c \log M$, plus order one 
(random) corrections. Therefore, all the fluctuations involved in the calculation of
$C_F$ will be of order $1$. The calculation of the average of the product of $\log Z$ 
for two different temperatures is a little more involved. At an intermediate level of
the computation, one finds, to order $b^0$:
\bea
\overline{\ell_1 \ell_2}&=&\Gamma^2(b)-\Gamma(b)\left[\frac{{\Gamma}({b}/{\mu_1})}
   {M^{{b}/{\mu_1}}\,\mu_1\,
     { {\Gamma}^{{b}/{\mu_1}}(1-\mu_1)  }}+ 1 \to 2 \right]\\
& &+\frac{\Gamma(b \beta)}{\mu_1 b M^{b \beta}}\left[\frac{1-b^2 \beta {\cal F}(\mu_1,\alpha)}
{\Gamma^{b \beta}(1-\mu_1)} + \alpha \frac{1-b^2 \beta {\cal F}(\mu_2,1/\alpha)}
{\Gamma^{b \beta}(1-\mu_2)}\right]  
\eea
with $\alpha=1+\varepsilon$, $\beta=(1+\alpha)/\mu_1$  and 
\be
{\cal F}(\mu,\alpha)=\int_0^1 \frac{dv}{v} \ \log\left[1+\frac{\mu}{\Gamma(1-\mu)} 
\int_0^\infty du \ \frac{\exp(-u)-\exp(-u^\alpha v)}{ u^{1+\mu}} \right] 
\ee
Expanding the previous result to order $\varepsilon^2$ and rearranging the terms finally
leads to $\kappa_F(\mu_1)$, which is too long to write explicitly here. Its dependence
in $\mu_1$ is given in Fig. 1. Of main interest is 
its behaviour for small temperatures and close to the transition point. For $\mu_1 \to 0$,
one finds that $\kappa_F$ goes to zero as:
\be
\kappa_F(\mu_1) \sim 0.905558... \mu_1^2.
\ee
For small temperatures, only the ground state contributes to the free-energy both
for $T_1$ and $T_2$. Hence, one indeed expects the sample to sample fluctuations
to be strongly correlated. For $\mu_1 \to 1$, on the contrary, one finds that $\kappa_F$ diverges:
\be
\kappa_F(\mu_1) \sim \frac{3 (4 \log 2 -1)}{{\pi }^2\,{\left( 1 - \mu_1 \right) }^3}
\sim \frac{0.538802...}{{\left( 1 - \mu_1 \right) }^3}
\ee
indicating that close to the `delocalisation' transition, the system tends to 
occupy rather different states when the temperature is slightly changed.
Finally, the correlation of free-energy fluctuations between $T_1 > T_c$
and $T_2 < T_c$ vanish as the size of the system tend to infinity.

 \begin{figure}
\hspace*{+1cm}\epsfig{file=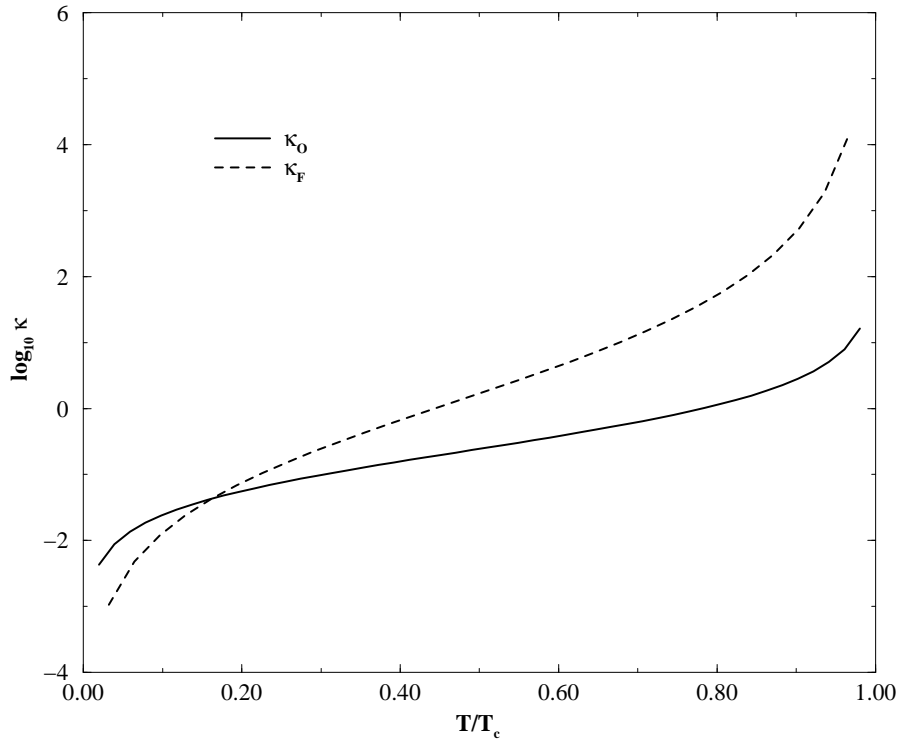,width=10cm,angle=270}
\vskip 0.3cm \caption{\small Plot of the rejuvenation susceptibilities $\kappa_O$ (plain line)
and $\kappa_F$ (dotted line) in semi-log scale, as a function of $\mu=T/T_c$. 
Note that both quantities
vanish at $T=0$ and diverge at $T=T_c$, with different exponents.
\label{fig1} }
\end{figure}

Therefore, except right at the transition, there is no `strong chaos' in the {\sc rem}, 
which would imply that the susceptibility $\kappa_F$ diverges with the size of the system. 
However, small temperature changes do lead to noticeable changes in the physical observables.
A perhaps more direct way to see this is to assign to each state $i$ 
a certain observable ${\cal O}_i$, independent the energy of this state. 
This can be for example, the position of the state
in space if the model describes (for example) the metastable states of a pinned 
interface, or the magnetisation of a state for spin system. We will assume for 
simplicity that $\sum_i {\cal O}_i = 0$. For a given set of random energies $\epsilon_i$, the
thermodynamical value of the observable is:
\be
\langle {\cal O} \rangle = \frac{1}{Z} \sum_{i=1}^M z_i {\cal O}_i
\ee
The average over disorder of this quantity is always zero. However, a fundamental 
difference arises between the case $T> T_c$ and $T<T_c$. In the former case 
and for large systems, $\langle{\cal O}\rangle$ tends to zero. This is related to the fact that 
the partition function is more or less evenly spread out on all states. On the contrary, 
for $T < T_c$, $\langle{\cal O}\rangle$ for a given system is a finite random quantity. Its 
variance is (for large $M$) given by:
\be
\overline{\langle {\cal O} \rangle^2}=(1-\mu) \sum_i {\cal O}_i^2 \qquad \mu=\frac{T}{T_c}
\ee   
Take for example the case where the observable ${\cal O}_i=x_i$ is the position of a 
particle in a box of size $2L$, $x_i= L(1-2i/M)$. In this case, the typical 
average position of the particle is $\langle x \rangle \sim \sqrt{1-\mu} L$ which scales
with the total size of the system, compared to $\langle x \rangle \sim L^{\zeta}$ with $\zeta < 1$ for 
$T>T_c$. Quenching the temperature from above $T_c$ therefore induces a complete rearrangement of 
the equilibrium properties of the system which occurs in a slow, aging way.

Now, let us how this observable changes when the temperature is slightly changed {\it within} the
glass phase. The calculation proceeds much as above, or as in \cite{Derrida3}. 
For small temperature shifts $\varepsilon \to 0$, we find:
\be
\overline{\left(\langle {\cal O} \rangle_1 -\langle {\cal O} \rangle_2\right)^2}
=\kappa_O(\mu_1) \varepsilon^2 \left(\sum_i {\cal O}_i^2\right),\label{susT}
\ee 
where the rejuvenation susceptibility is given by a very lengthy expression. 
This quantity is plotted in Fig. 1. Its small temperature behaviour is given by:
\be
\kappa_O(\mu_1) \sim \frac{12 + \pi^2}{18} \,\mu_1 = 1.21497802... \mu_1,
\ee
where its divergence for $\mu_1 \to 1$ is given by:
\be
\kappa_O(\mu_1) \sim  \frac{1}{3(1-\mu_1)}.
\ee 
Therefore, we again find a divergence of the rejuvenation susceptibility near $T_c$.
For $T > T_c$, we find that $\kappa_O$ vanishes as a ($T$ dependent) power of $M$.

We then find a very interesting situation: Eq. (\ref{susT}) tells us that 
when the temperature is slighlty changed, the mean position of the particle (say) has 
to evolve by an amount of order $\sqrt{\kappa} \varepsilon L$ that is {\it proportional to the 
size of the system} (and with a diverging amplitude when $T \to T_c$). 
In this sense, rejuvenation is strong since a small temperature 
change will induce a rather large response of the system. However, since the number of
states occupied by the particle remains finite in the whole glassy phase $T < T_c$, the probability
to find the system in the same state at $T_1$ and $T_2$ remains finite in the limit of 
large systems \cite{Compte/Bouchaud}. This probability is directly related to $\kappa_O$,
and reads:
\be
P_{12} = 1 - \frac{\mu_1 + \mu_2}{2} - \kappa_O \varepsilon^2. 
\ee 
(Note that $P_{11}=1-\mu_1$ as it should \cite{Derrida3,Bouchaud/Mezard}).
In the p-spin glass model where two states are generically orthogonal, the resulting 
two temperature overlap function is therefore given by:
\be
\overline{P(q,T_1,T_2)} = (1-P_{12}) \delta(q) + P_{12} \ \delta(q-q(T_1,T_2)).
\ee

In summary, we have shown that the Random Energy Model has interesting rejuvenation properties
in its frozen phase. Different `susceptibilities' to temperature changes, for the free-energy and
for other (magnetic) observables, can be computed exactly. These susceptibilities diverge at the 
transition temperature with different exponents. Since the {\sc rem} seems to be relevant to
many physical situations, the mechanism found here is probably of broad interest.  However, 
the coexistence of rejuvenation and memory seen in the spin-glass 
experiments cannot be accounted for by the simplest version of the {\sc rem}, because the
evolution at $T_2$ will have a significant influence on the properties measured at $T_1$ 
after reheating. One can generalize the {\sc rem} along the lines of \cite{Derrida/Gardner} by 
allowing a hierarchy of phase transitions $T_{c,n}$. As argued in \cite{Bouchaud/Dean} and numerically
demonstrated in \cite{Nemoto/Sasaki}, each crossing of a phase transition $T_{c,n}$ induces a strong 
rejuvenation signal (much as calculated here) with a slow aging dynamics for $T < T_{c,n}$
and a fast return to equilibrium for $T > T_{c,n}$, accounting for the memory effect.

{\it Acknowledgements} 
M. S. thanks the Spanish Ministry of Education and Culture for a visit grant to
Saclay. 
J.P. B. wants to thank D. S. Fisher and Harvard University for hospitality during the period this
work was completed. We are indebted to D. S. Fisher, M. M\'ezard, F. Ritort, J. Sethna, R. da Silveira and L.H. Tang for interesting remarks and discussions.

\end{document}